%Paper: gr-qc/9511007
%From: hayward@murasaki.scphys.kyoto-u.ac.jp
%Date: Thu, 2 Nov 95 16:51:11 JST

%%%%%%%%%%%%%%%%%%%%%%%%%%%%%%%%%%%%%%%%%%%%%%%%%%%%%%%%%%%%%%%%%%%%%%%%%%%%%%
% Complex lapse, complex action and path integrals (Sean A. Hayward)
%%%%%%%%%%%%%%%%%%%%%%%%%%%%%%%%%%%%%%%%%%%%%%%%%%%%%%%%%%%%%%%%%%%%%%%%%%%%%%

\font\mbf=cmbx10 scaled\magstep1
\font\sm=cmr7

\def\bs{\bigskip}
\def\ms{\medskip}
\def\np{\vfill\eject}

\def\ni{\noindent}
\def\cl{\centerline}

\def\title#1{\cl{\mbf #1}}
\def\ref#1#2#3#4{#1\ {\it#2\ }{\bf#3\ }#4\par}
\def\refb#1#2#3{#1\ {\it#2\ }#3\par}
\def\CQG{Class.\ Qu.\ Grav.}
\def\PL{Phys.\ Lett.}
\def\PR{Phys.\ Rev.}

\def\d{\hbox{d}}

\def\p{\partial}
\def\L{{\cal L}}
\def\half{{\textstyle{1\over2}}}
\def\th{{\textstyle{3\over2}}}

\magnification=\magstep1

\title{Complex lapse, complex action and path integrals}
\bs\cl{\bf Sean A. Hayward}
\ms\cl{Department of Physics, Kyoto University, Kyoto 606-01, Japan}
\ms\cl{{\tt hayward@murasaki.scphys.kyoto-u.ac.jp}}
\bs\ni
{\bf Abstract.}
Imaginary time is often used in quantum tunnelling calculations.
This article advocates a conceptually sounder alternative: complex lapse.
In the ``3+1'' action for the Einstein gravitational field
minimally coupled to a Klein-Gordon field,
allowing the lapse function to be complex yields a complex action
which generates both the usual Lorentzian theory and its Riemannian analogue,
and in particular allows a change of signature between the two.
The action and variational equations are manifestly well defined
in the Hamiltonian representation,
with the momentum fields consequently being complex.
The complex action interpolates between the Lorentzian and Riemannian actions
as they appear formally in the respective path integrals.
Thus the complex-lapse theory provides
a unified basis for a path-integral quantum theory of gravity
involving both Lorentzian and Riemannian aspects.
A major motivation is the quantum-tunnelling scenario
for the origin of the universe.
Taken as an explanation for the observed quantum tunnelling of particles,
the complex-lapse theory determines that
the argument of the lapse for the universe now is extremely small but negative.
\bs\cl{PACS: 04.20.Fy, 04.60.Gw, 98.80.Hw}
\bs\cl{16th October 1995}
\np\ni
{\bf I. Introduction}
\ms\ni
Considerable interest is aroused by the possibility of
describing the origin of the universe by quantum tunnelling;
consult, for instance, the review by Hawking [1].
The mathematics describing the quantum tunnelling of a particle
out of a classically impenetrable region,
when applied to the universe as a whole,
may be interpreted as describing the quantum tunnelling of the universe
from a state of Riemannian signature
to the currently observed Lorentzian signature.
In the simplest model,
an inflationary de Sitter universe tunnels out of a Riemannian 4-sphere [1].
The non-singular nature of this model is appealing in comparison to
the classical model of the origin of the universe
as a singularity of mysterious provenance.
Such considerations have led to the idea that
a quantum theory of gravity should involve imaginary time,
as expressed, for instance, in the Hawking no-boundary proposal,
in which the fundamental object is a wave function defined formally
as a path integral over compact Riemannian 4-manifolds [1].

Of particular interest are real tunnelling solutions,
described by Gibbons \& Hartle [2] as consisting of
a Riemannian manifold and a Lorentzian manifold
joined at a common boundary of vanishing extrinsic curvature.
The simple model mentioned above can be interpreted in this sense,
consisting of a Riemannian 4-hemisphere
joined to the expanding half of the de Sitter universe.
Moreover, this model has the highest probability
among the class of real tunnelling solutions
to the Einstein equations with positive cosmological constant
and compact Riemannian part [2],
where the relative probability is determined by
the value of the Riemannian action.
Thus there is the prediction---for this matter model---that
the most probable early universe
is the inflationary, maximally symmetric de~Sitter universe [3].
This prediction agrees well with observations of cosmological isotropy
and current inflationary-universe theory [4,5].
The observed small anisotropy might be explained by
quantum fluctuations about the classical solution [1,4,5].

Accepting the physical relevance of such real tunnelling solutions,
there remains the question of how to mathematically describe them
and how to generalise to complex tunnelling solutions.
Sometimes the solutions are described in terms of
the Wick rotation $t\to it$ of the time coordinate $t$.
The above geometrical description is essentially that of Gibbons \& Hartle [2],
but there has been some question about
the geometrical meaning of joining the two manifolds.
The key observation is that a real tunnelling solution
describes a change of signature of the space-time metric,
and therefore that the relevant topic
is the Einstein equation for signature-changing metrics,
as far as this makes sense.
Adopting a ``3+1'' decomposition of space-time
based on the hypersurface of signature change,
this is described by the lapse function $l$ being imaginary
in the Riemannian region and real, as usual, in the Lorentzian region.
Ellis et al.\ [6] effected this by using the squared lapse funtion $\nu=l^2$
and allowing $\nu$ to be negative.
Unfortunately,
Ellis et al.\ derived so-called solutions to the Einstein equations
with non-vanishing extrinsic curvature at the junction,
which actually do not satisfy the Einstein equations.
This has been corrected [7].
Ellis et al.\ also argued that discontinuous $\nu$ was preferable
to smooth $\nu$, but the two choices actually give equivalent solutions.
Either approach, formulated correctly,
describes a change of signature subject to the Einstein equations.
Similarly, one may use the inverse squared lapse function $\lambda=l^{-2}$ [8].
However, since $\nu$ or $\lambda$ is assumed to be real,
these approaches are restricted to real tunnelling solutions.

A more general approach
is simply to allow the lapse function $l$ to be complex.
This is the approach explored in this article.
A similar suggestion in a simpler context has been made by Greensite [9,10],
who concluded that Lorentzian signature is dynamically preferred.
It is worth repeating Greensite's point that
this approach differs from allowing the time coordinate $t$ to be imaginary,
which is the approach of Euclidean quantum theory
and consequently the usual approach in quantum cosmology.
So the complex-lapse theory developed here is to be regarded as
an alternative to the conventional imaginary-time theory.
In the closing Section V it will be argued that
the complex-lapse theory clarifies
some aspects of quantum theory, particularly quantum tunnelling,
that were difficult to interpret in the imaginary-time theory.

In order to have a simple but non-trivial matter field,
a real Klein-Gordon field will be taken,
minimally coupled to the Einstein field.
Section II reviews the standard ``3+1'' decomposition of space-time
and the Lagrangian and Hamiltonian forms of the dynamics.
In Section III, the generalisation to complex lapse is effected.
There is a subtle difference
between the Lagrangian and Hamiltonian representations,
since the Lagrange transformation is singular
where the lapse function vanishes.
The Hamiltonian theory is preferred since it is non-singular.
This entails certain restrictions on the Lagrangian theory,
which render it well defined.
In Section IV,
it is observed that the formal path integral of the complex-lapse theory
interpolates between the usual Lorentzian and Riemannian path integrals.
Again, this generalises an observation of Greensite [9,10].

The complex action compares favourably with
the actions suggested by Embacher [11]
to describe a real signature change in the vacuum Einstein theory.
Embacher's actions are all real, so do not give the correct path integrals.
Moreover, for Embacher's preferred action,
the variational equations differ from the Einstein equations,
i.e.\ the vanishing of the Einstein tensor as calculated from the metric.
Embacher's aim was to show that the equations to which the so-called solutions
of Ellis et al.\ [6] are actually solutions---the Einstein equations
away from the signature transition---can be derived from an action.
This may be so, but the relevance is questionable.
\bs\ni
{\bf II. Real lapse}
\ms\ni
The action for the Einstein-Klein-Gordon field is given by
$$S=\int_M\hat{*}\left(\hat{R}-\half g^{-1}(\d\phi,\d\phi)-V\right)\eqno(1)$$
with units $16\pi G=1$,
where $M$ is a 4-manifold with exterior derivative $\d$,
$g$ is the space-time metric on $M$,
with volume form $\hat{*}$ and Ricci scalar $\hat{R}$,
and $\phi$ is the real Klein-Gordon field on $M$, with potential $V(\phi)$.

A ``3+1'' decomposition [12] may be made either globally if
$M\cong T\times\Sigma$,
or locally by taking an embedding from $T\times\Sigma$ to $M$,
where $T$ is a real interval and $\Sigma$ a 3-manifold.
When taking variations it is easiest to assume that $\Sigma$ is compact,
but non-compact $\Sigma$ can be treated by demanding that
the fields have certain asymptotic behaviour.
Denote the coordinates by $t\in T$ and $x\in\Sigma$.
Define the evolution vector $v=\p/\p t$, meaning that $v(\d t)=1$.
Define the lapse function $l$ by $l^{-2}=-g^{-1}(\d t,\d t)$,
the spatial 3-metric $h=g+l^{-2}\d t\otimes\d t$
and the shift 3-vector $s=h^{-1}h(v)$.
Then in a basis $(\p/\p t,\p/\p x)$,
the metric takes the form
$$g=\pmatrix{h(s,s)-l^2&h(s)\cr h(s)&h\cr}.\eqno(2)$$
Take configuration fields $q=(\phi,h,s,l)$
and the corresponding velocity fields $\L_v q$,
where $\L_u$ denotes the Lie derivative along a vector $u$.
Now write the action (1) in the Lagrangian form
$$S=\int_T \d t \int_\Sigma L(q,\L_v q)\eqno(3)$$
by removing second time derivatives in the form of total divergences.
The Lagrangian turns out to be [7]
$$L={*}\half l^{-1}\left(\half(\L_{v-s}h)^\sharp:\L_{v-s}h
-\half(h^\sharp:\L_{v-s}h)^2+(\L_{v-s}\phi)^2\right)
+{*}l\left(R-\half D\phi\cdot D^\sharp\phi-V\right)\eqno(4)$$
where $h$ has volume form $*$, covariant derivative $D$ and Ricci scalar $R$,
a dot ($\cdot$) denotes a single symmetric contraction,
a colon (:) denotes a double symmetric contraction,
and the sharp ($\sharp$) denotes the contravariant dual
with respect to the inverse spatial metric $h^{-1}=h^\sharp$.\footnote\dag
{{\sm That is, index raising.
The index notation has been avoided because
it encourages unwitting raising or lowering of indices,
which would be confusing when taking variations
in which the metric is a variable.
The index notation also allows easy confusion between
expressions which appear identical except for the index labelling,
but have quite different meanings.}}

To be more mathematically precise,
the basic object is a fibre bundle $Q$ over $\Sigma$, the configuration bundle,
with smooth sections $q\in C^\infty Q$.
The velocity fields live in the tangent bundle $TC^\infty Q$,
viz.\ $(q,\L_vq)\in TC^\infty Q$.
The Lagrangian is a map $L:TC^\infty Q\to F$,
where $F$ denotes the space of 3-forms on $\Sigma$.
Variations may be defined by
$$\delta q=\left.{\d\over{\d\alpha}}\tilde
q(\alpha)\right|_{\alpha=0}\eqno(5)$$
where $\tilde q(\alpha)$ is a one-parameter family with $\tilde q(0)=q$.
Functional derivatives are defined so as to satisfy the chain rule, e.g.
$$\delta\Phi={\delta\Phi\over{\delta q}}\delta q\eqno(6)$$
for a functional $\Phi(q)$,
where the functional derivative $\delta\Phi/\delta q$
is a linear map from $C^\infty Q$ to the space where $\Phi$ lives.
Particularly important is the Lagrange transformation
from velocity fields $(q,\L_vq)\in TC^\infty Q$
to momentum fields $(q,\delta L/\delta\L_vq)\in T^*C^\infty Q$,
where $T^*C^\infty Q$ denotes the cotangent bundle.
The Hamiltonian is a map $H:T^*C^\infty Q\to F$ defined by
$$H={\delta L\over{\delta\L_vq}}\L_vq-L\eqno(7)$$
where the inverse Lagrange transformation is implicit.

The principle of stationary action reads
$$0=\delta S=\int_T\d t\int_\Sigma\left({\delta L\over{\delta q}}\delta q
+{\delta L\over{\delta\L_v q}}\delta\L_v q\right)
=\int_T\d t\int_\Sigma\left({\delta L\over{\delta q}}
-\L_v\left({\delta L\over{\delta\L_v q}}\right)\right)\delta q\eqno(8)$$
for variations $\delta q$ which vanish on $\p T$.
This yields the Euler-Lagrange equations in the integral form
$$\int_\Sigma\left({\delta L\over{\delta q}}
-\L_v\left({\delta L\over{\delta\L_v q}}\right)\right)=0.\eqno(9)$$
The explicit form of these equations
is equivalent to the more convenient Hamilton equations derived below.

Perform the Lagrange transformation from the velocity fields $\L_v q$
to the momentum fields $\hat q=\delta L/\delta\L_v q$.
The independent momentum fields are
$$\eqalignno
{&\hat\phi={\delta L\over{\delta\L_v \phi}}
={*}l^{-1}\L_{v-s}\phi&(10a)\cr
&\hat h={\delta L\over{\delta\L_v h}}
={*}\half
l^{-1}\left(\L_{v-s}h-(h^\sharp:\L_{v-s}h)h\right)^\sharp.&(10b)\cr}$$
Since the Lagrangian is independent of $\L_v l$ and $\L_v s$,
the corresponding momentum fields vanish and the Hamiltonian is defined by
$$H(\phi,h,s,l,\hat\phi,\hat h)=\hat\phi\L_v\phi+\hat h:\L_v h-L\eqno(11)$$
where the inverse Lagrange transformation is implicit.
The variational principle leads to the Hamilton equations in the integral form
$$\eqalignno
{&\int_\Sigma{*}\left(\L_v\phi-{\delta
H\over{\delta\hat\phi}}\right)=0&(12a)\cr
&\int_\Sigma{*}\left(\L_v h-{\delta H\over{\delta\hat h}}\right)=0&(12b)\cr
&\int_\Sigma\left(\L_v\hat\phi+{\delta H\over{\delta\phi}}\right)=0&(12c)\cr
&\int_\Sigma\left(\L_v\hat h+{\delta H\over{\delta h}}\right)=0&(12d)\cr
&\int_\Sigma{\delta H\over{\delta s}}=0&(12e)\cr
&\int_\Sigma{\delta H\over{\delta l}}=0.&(12f)\cr}$$
Explicitly, the Hamiltonian (11) is
$$H={l\over*}\left(\hat h:\hat h^\flat-\half(h:\hat h)^2+\half\hat\phi^2\right)
+{*}l\left(-R+\half D\phi\cdot D^\sharp\phi+V\right)
+\hat h:\L_s h+\hat\phi\L_s\phi\eqno(13)$$
where the flat ($\flat$) denotes the covariant dual with respect to $h$.
Most of the variations involved in the Hamilton equations (12)
are straightforward, the non-trivial one being
$$\delta\int_\Sigma{*}lR=\int_\Sigma{*}R\delta l
+\int_\Sigma{*}\left(-lG+D\otimes Dl-hD^2l\right)^\sharp:\delta h\eqno(14)$$
where $G$ is the Einstein tensor of $h$ and $D^2=D\cdot D^\sharp$.
When evaluating the variations,
one may remove total divergences $\int_\Sigma{*}D\cdot u=0$
for vectors $u$ on $\Sigma$,
including those of the form $\int_\Sigma\L_s({*}f)=\int_\Sigma{*}D\cdot(fs)$
for functions $f$ on $\Sigma$.
Noting also the identity
$$\delta({*}1)={*}\half h^\sharp:\delta h\eqno(15)$$
the variations appearing in (12) read explicitly
$$\eqalignno
{&\int_\Sigma{*}{\delta H\over{\delta\hat\phi}}
=\int_\Sigma\left(l\hat\phi+{*}\L_s\phi\right)&(16a)\cr
&\int_\Sigma{*}{\delta H\over{\delta\hat h}}
=\int_\Sigma\left(l\left(2\hat h^\flat-(h:\hat h)h\right)
+{*}\L_s h\right)&(16b)\cr
&\int_\Sigma{\delta H\over{\delta\phi}}
=\int_\Sigma\left({*}l{\p V\over{\p\phi}}-{*}D\cdot(l D^\sharp\phi)
-\L_s\hat\phi\right)&(16c)\cr
&\int_\Sigma{\delta H\over{\delta h}}
=\int_\Sigma{l\over*}\left(
2\hat h\cdot h\cdot\hat h-(h:\hat h)\hat h-\half\left(
\hat h:\hat h^\flat-\half(h:\hat h)^2+\half\hat\phi^2\right)h^\sharp\right)\cr
&\qquad\qquad+\int_\Sigma{*}\half l\left(
-D\phi\otimes D\phi+\left(
\half D\phi\cdot D^\sharp\phi+V\right)h\right)^\sharp\cr
&\qquad\qquad+\int_\Sigma{*}\left(lG-D\otimes Dl+hD^2l\right)^\sharp
-\int_\Sigma\L_s\hat h&(16d)\cr
&\int_\Sigma{\delta H\over{\delta s}}
=\int_\Sigma\left(\hat\phi D\phi-2(D\cdot\hat h)^\flat\right)&(16e)\cr
&\int_\Sigma{\delta H\over{\delta l}}
=\int_\Sigma{1\over*}
\left(\hat h:\hat h^\flat-\half(h:\hat h)^2+\half\hat\phi^2\right)
+\int_\Sigma{*}\left(-R+\half D\phi\cdot D^\sharp\phi+V\right).&(16f)\cr}$$
It is convenient to introduce alternative momentum fields in the form of
a function $\psi$ and a contravariant metric $p$ defined by
$$\eqalignno
{&\psi={\hat\phi\over*}
=l^{-1}\L_{v-s}\phi&(17a)\cr
&p={\hat h\over*}
=\half l^{-1}\left(\L_{v-s}h-(h^\sharp:\L_{v-s}h)h\right)^\sharp.&(17b)\cr}$$
Using the identity
$$\L_{v-s}({*}1)={*}\half h^\sharp:\L_{v-s}h=-{*}\half lh:p\eqno(18)$$
and the variations (16),
the Hamilton equations (12) integrate explicitly to
$$\eqalignno
{&\L_{v-s}\phi=l\psi&(19a)\cr
&\L_{v-s}h=l\left(2p^\flat-(h:p)h\right)&(19b)\cr
&\L_{v-s}\psi=l\left(\half(h:p)\psi+D^2\phi-\p V/\p\phi\right)
+Dl\cdot D^\sharp\phi&(19c)\cr
&\L_{v-s}p=l\left(-2p\cdot h\cdot p+\th(h:p)p
+\half(D\phi\otimes D\phi)^\sharp-G^\sharp\right)\cr
&\qquad\qquad+\half l\left(p:p^\flat-\half(h:p)^2 +\half\psi^2
-\half D\phi\cdot D^\sharp\phi-V\right)h^\sharp\cr
&\qquad\qquad+\left(D\otimes Dl-hD^2l\right)^\sharp&(19d)\cr
&0=\psi D\phi-2(D\cdot p)^\flat&(19e)\cr
&0=p:p^\flat-\half(h:p)^2+\half\psi^2
-R+\half D\phi\cdot D^\sharp\phi+V.&(19f)\cr}$$
These are the Einstein-Klein-Gordon equations in Hamiltonian form.
\bs\ni
{\bf III. Complex lapse}
\ms\ni
Consider generalising the dynamical theory of the previous section
by allowing the lapse function $l$ to be complex.
Inspecting the metric (2) reveals two special cases:
purely real and purely imaginary $l$,
which respectively describe metrics of Lorentzian and Riemannian signature.
In the Lorentzian case, one has the usual Einstein-Klein-Gordon theory,
while in the Riemannian case,
one has the corresponding Riemannian (often called Euclidean) theory.
In both cases, one can check that the equations are correct
by using (19a) and (19b) to eliminate the momenta $(\psi,p)$:
(19c) yields the Klein-Gordon equation
and (19d), (19e) and (19f) yield respectively
the stress, momentum and energy components of the Einstein equation.

In particular, there is the possibility of a signature transition
between Lorentzian and Riemannian states,
referred to in quantum cosmology as real tunnelling [2].
There is a slight problem with this scenario
in the Lagrangian (or velocity) representation:
if, as is normally assumed, the lapse $l$ is continuous,
then it must vanish at a signature transition.
But the kinetic term in the Lagrangian (4)
is linear in the inverse lapse $l^{-1}$.
Thus if the lapse $l$ vanishes, the Lagrangian becomes singular.
Similarly, the Euler-Lagrange equations (9) contain negative powers of $l$,
so they also become singular if $l$ vanishes.
These negative powers of $l$
have been the cause of a great deal of confusion in the literature.
Fortunately,
this problem does not arise in the Hamiltonian (or momentum) representation.
The Hamiltonian (13) is manifestly non-singular; it contains $l$ only linearly.
Similarly, the Hamilton equations (19) are manifestly non-singular;
they contain no negative powers of $l$.
The trick is that the Lagrange transformation (10) is singular,
being linear in the inverse lapse.
Thus if the momentum fields are well behaved at the signature transition,
then the velocity fields must be better behaved.

This suggests adopting the Hamiltonian representation,
which is anyway dynamically natural,
whether in quantum theory or classical theory.
Specifically, the proposed theory is as follows:
{\it take the usual Hamiltonian theory, but allow
the lapse $l$ and momenta $(\hat\phi,\hat h)$ or $(\psi,p)$ to be complex.}
In particular, the other configuration fields $(\phi,h,s)$ remain real,
as do the coordinates $(t,x)$.
The physical fields $(\phi,h,s,l,\psi,p)$
are assumed to be continuous, with sufficient differentiability in $t$ and $x$
that the Hamilton equations (19) are well defined and admit solutions.
It follows from the inverse Lagrange transformation,
or equivalently from the Hamilton equations (19a) and (19b), that
$$\L_{v-s}(\phi,h)=O(l)\quad\hbox{as $l\to0$.}\eqno(20)$$
This renders the Lagrangian (4) and action (3) well defined.
Actually,
all the formulas of the previous section are well defined in the new context.
The asymptotic behaviour (20) should be understood
in subsequent references to the Lagrangian representation;
it is to be regarded as
just a representation of the underlying Hamiltonian theory,
determined by the (non-singular) inverse Lagrange transformation.

A simple consequence of allowing complex lapse is that
the momentum fields $(\psi,p)$ must also be generally complex
if the field equations are to be satisfied.
In fact, the Lagrange transformation (17) shows that
the arguments are synchronised:
$$\arg\psi=\arg p=-\arg l\qquad\hbox{(mod $\pi$).}\eqno(21)$$
Thus the momentum fields $(\psi,p)$ are purely real in Lorentzian regions
and purely imaginary in Riemannian regions.
In particular, at a hypersurface $\Sigma_0$ where the signature changes
from Riemannian to Lorentzian, continuity requires
$$l|_{\Sigma_0}=\psi|_{\Sigma_0}=p|_{\Sigma_0}=0.\eqno(22)$$
These are the junction conditions for real signature change
(or real tunnelling).
In particular, the momentum fields vanish at a signature transition.
These conditions are well known in the context of quantum cosmology [2]
and can be derived independently using several different representations [7,8].
It is these junction conditions that are not satisfied by
the so-called solutions of Ellis et al.\ [6].

Note that in a quantum version of the theory,
the arguments of the momenta and lapse need not be synchronised as in (21);
this condition indicates when a purely classical solution is possible.
Here ``classical'' means non-quantum, not non-Lorentzian.
\bs\ni
{\bf IV. Path integrals}
\ms\ni
The Feynman path integral [13] for the Einstein-Klein-Gordon field
may be written formally as
$$Z=\sum_\Sigma\int_{C^\infty Q}\star\exp(iS)\eqno(23)$$
with units $\hbar=1$,
where $S$ is the action (3--4) in Lagrangian form
and $\star$ is a suitable measure
on the configuration fields $q\in C^\infty Q$,
including a weighting for the topology of $\Sigma$.
The ambiguity in $\star$ is the reason for the qualifier ``formally''.
It should be understood that $Z$ is a functional of $(q_1,q_0)$,
the final and initial values of $q$
at the boundaries $t=t_1$ and $t=t_0$ of $T$,
enforced by projections $\chi(q_1-q(t_1))\chi(q_0-q(t_0))$ in $\star$.
Here the characteristic function $\chi$ vanishes except for $\chi(0)=1$.

Under the generalisation to complex lapse function $l$
described in the preceding section, the action $S$ becomes complex.
In particular, for negative imaginary $l$, $-iS$ is real
and coincides with the Riemannian (often called Euclidean) action $I$ [1].
Thus the Riemannian action may also be generalised to the complex-lapse theory
by the definition
$$I=-iS.\eqno(24)$$
This allows the path integral (23) to be written formally
in the Riemannian (or Euclidean) form [1]
$$Z=\sum_\Sigma\int_{C^\infty Q}\star\exp(-I).\eqno(25)$$
To check the above statements,
note that the action may be written in kinetic-potential form:
$$S=\int_T\d t\int_\Sigma(l^{-1}K-lU)\eqno(26)$$
where the kinetic energy $K$ and potential energy $U$ are given by
$$\eqalignno
{&K={*}\half\left((\L_{v-s}\phi)^2
+\half(\L_{v-s}h)^\sharp:\L_{v-s}h-\half(h^\sharp:\L_{v-s}h)^2\right)&(27a)\cr
&U={*}\left(V+\half D\phi\cdot D^\sharp\phi-R\right).&(27b)\cr}$$
Taking normal coordinates for the Lorentzian and Riemannian theories
separately,
the appropriate actions reduce respectively to the expected forms:
$$\eqalignno
{&S\Big\vert_{l=1}=\int_T\d t\int_\Sigma(K-U)&(28a)\cr
&I\Big\vert_{l=-i}=\int_T\d t\int_\Sigma(K+U).&(28b)\cr}$$
The actions (28a) and (28b) and the path integals (23) and (25)
are the standard expressions
in the respective Lorentzian [13] and Riemannian [1] theories,
with all signs correct.

An equivalent formulation involves the {\it scaling\/} (or Riemannian lapse)
$$\ell=il\eqno(29)$$
instead of $l$.
Then the Riemannian action may be written
$$I=\int_T\d t\int_\Sigma(\ell^{-1}K+\ell U)\eqno(30)$$
and the Lorentzian and Riemannian normal coordinates are given respectively by
$\ell=i$ and $\ell=1$.
The transformation $l\to\ell=il$ is similar to the Wick rotation $t\to it$,
but for lapse instead of time.
Note that this rotation has the sign standard in Euclidean quantum theory [1],
not the opposite sign sometimes suggested [5,9].

Two points should be emphasised to distinguish the above proposal
from traditional wisdom.
Firstly, it is the lapse $l$ rather than time $t$ which is complex.
Secondly, the proposal is not to use the rotation (29) to convert
a purely Lorentzian theory with real $l$
to a purely Riemannian theory with real $\ell$,
but to allow $l$ (or $\ell$) to be fully complex in a general theory
which includes the Lorentzian and Riemannian theories as special cases.
Both differences are summarised in the phrase ``complex lapse''
versus ``imaginary time''.

In summary, in the complex-lapse theory,
either the usual action $S$ or the Riemannian action $I$
generalises to provide a path integral which interpolates between
the Lorentzian and Riemannian path integrals,
given respectively by positive real $l$ (or positive imaginary $\ell$)
and negative imaginary $l$ (or positive real $\ell$).
This is true not just for the Einstein-Klein-Gordon field
but for any field whose action takes the kinetic-potential form (26),
where $K$ is quadratic in the velocity fields
and $U$ is independent of the velocity fields.
This reveals that the Lorentzian and Riemannian actions
are not as distinct as they appear in conventional quantum theory,
but are two facets of a single complex action
involving a complex lapse function.

If the complex-lapse theory is to be applied,
quantisation still needs to be made concrete,
for instance by properly defining the measure $\star$
in the path integral (23) or (25).
At least,
this could be tried in a highly symmetric effective (minisuperspace) model.
Here just one point will be made concerning the measure, as follows.
One wishes to integrate only over the physical fields,
factoring out coordinate transformations which generate equivalent solutions.
In particular, the shift vector $s$ concerns the choice of evolution direction,
so need not be integrated over.
Similarly, the modulus of the lapse function $l$
just concerns the scaling of the coordinate $t$,
so need not be integrated over.
But the argument of the lapse is physically distinct
and should be integrated over.
So the natural choice for measure in the complex plane of the lapse
is to integrate over a circle centred on the origin.
By Cauchy's residue theorem,
this integral will just pick up the $l^{-1}$ term in the integrand.
This procedure is more specific than that of the imaginary-time theory,
where there is an ambiguity in the choice of contour [1].

Similarly,
decoherence functionals [14] also generalise to the complex-lapse theory.
\bs\ni
{\bf V. Discussion}
\ms\ni
The main point of this article is that
allowing the lapse function to be complex
in the ``3+1'' action for the Einstein-Klein-Gordon field
yields a complex action
which generates both the usual Lorentzian theory and its Riemannian analogue,
and in particular allows a change of signature between the two.
The action and variational equations are manifestly well defined
in the Hamiltonian representation,
with the momentum fields consequently being complex.
Moreover,
the complex action interpolates between the Lorentzian and Riemannian actions
as they appear formally in the respective path integrals.
So the complex action provides a unified basis
for the Einstein-Klein-Gordon field
under both Lorentzian and Riemannian signatures.
This also suggests that the complex action provides
a natural path-integral basis for a quantum theory
which allows a general treatment of quantum tunnelling.
Of particular interest is the possible origin of the universe
as a quantum tunnelling process.

To a large extent this supports the belief that
a quantum theory of gravity should involve something like imaginary time [1].
But note that it is not the time coordinate $t$ that is complex,
but the lapse function $l$; the time $t$ remains real.
In comparison,
calculations in Euclidean quantum gravity involve complex time explicitly.
It should be emphasised that
the imaginary-time and complex-lapse theories are fundamentally different:
in one, the space-time dimension expands by one, while in the other,
space-time stays intact and one of the physical fields becomes complex.
Since the latter is a less radical revision of conventional physical concepts,
it seems worth exploring.

The main reason for believing in imaginary time seems to be
its role in Euclidean quantum theory for flat space-time.
Here one Wick-rotates the time coordinate $t\to it$,
performs some calculations and then Wick-rotates back to interpret the results.
While this works in flat space-time,
it is not clear that it gives physically correct answers in curved space-time.
This procedure is reminiscent of
the imaginary-time approach to classical special relativity,
where instead of the Minkowski metric with real time,
one works with a Euclidean metric and imaginary time.
This imaginary-time approach works in special relativity
but not in general relativity;
the signature is a property of the metric, not the coordinates.
So it may be argued that the use of imaginary time
is no more appropriate in quantum gravity than in classical gravity;
variable or complex signature must be encoded in the metric,
not the coordinates.

A great conceptual problem with the imaginary-time theory is that
one has to imagine an additional dimension of time.
This imaginary dimension is somehow inaccessible classically,
yet responsible for quantum tunnelling effects.
If one really believes in this imaginary time,
one would have to explain why it is not directly observed.
Alternatively, if imaginary time is just a mathematical trick
for performing quantum-tunnelling calculations,
one is still left wondering what the underlying physical reality is.

Allowing complex lapse seems conceptually clearer:
the lapse is a complex physical field which may have any value,
and the universe now happens to have almost real lapse.
But the lapse is not exactly real;
the imaginary part of the lapse contributes to path integrals.
This suggests an alternative description of quantum tunnelling,
including the observed quantum tunnelling of particles.
Consistency with such observation requires
the argument of the lapse now to be extremely small but negative.
This sign is the physical basis for the conventional sign of the Wick rotation,
which otherwise seems arbitrary.
So a radical new idea has arisen:
the physical basis for quantum tunnelling is that
the lapse function of the universe is complex.
If so, then assuming reality of the other components of the metric
loses its original justification.

Once complex lapse is deemed possible, it must be accepted as the general rule.
Thus the current special state of the universe requires some explanation.
The argument of Greensite [9,10] that
Lorentzian signature is preferred may be relevant here.
In contrast, the early universe at the quantum cosmology stage
need not have had anything like real lapse.
Similarly, the gravitational momentum $p$ is now almost real,
but need not have been originally.
Indeed, Hawking's no-boundary proposal [1] could be re-interpreted
as implying that the universe was originally dominated by imaginary lapse.
So to explain the origin of the observed universe by quantum tunnelling,
one might look for some dynamical mechanism,
perhaps with special initial or boundary conditions,
driving the universe towards its present state
with the lapse and gravitational momentum being almost real.
It is tempting to speculate on the possible role of inflation here.
\bs\ni
Research supported by the Japan Society for the Promotion of Science.
\bs
\begingroup
\parindent=0pt\everypar={\global\hangindent=20pt\hangafter=1}\par
{\bf References}\par
\refb{[1] Hawking S W 1987 in}{Three Hundred Years of Gravitation}
{ed: Hawking~S~W \& Israel~W (Cambridge University Press)}
\ref{[2] Gibbons G W \& Hartle J B 1990}\PR{D42}{2458}
\ref{[3] Hayward S A 1993}\CQG{10}{L7}
\refb{[4] Blau S K \& Guth A H 1987 in}{Three Hundred Years of Gravitation}
{ed: Hawking~S~W \& Israel~W (Cambridge University Press)}
\refb{[5] Linde A 1987 in}{Three Hundred Years of Gravitation}
{ed: Hawking~S~W \& Israel~W (Cambridge University Press)}
\ref{[6] Ellis G, Sumeruk A, Coule D \& Hellaby C 1992}\CQG{9}{1535}
\ref{[7] Hayward S A 1992}\CQG9{1851; erratum 2453}
\refb{[8] Hayward S A 1995}
{Signature change at material layers and step potentials}{gr-qc/9509052}
\ref{[9] Greensite J 1993}\PL{B300}{34}
\ref{[10] Carlini A \& Greensite J 1993}\PR{D49}{866}
\ref{[11] Embacher F 1995}\PR{D51}{6474}
\refb{[12] Fischer A E \& Marsden J E 1979 in}
{General Relativity, an Einstein Centenary Survey}
{ed: Hawking~S~W \& Israel~W (Cambridge University Press)}
\refb{[13] Itzykson C \& Zuber J-B 1980}{Quantum Field Theory}{(McGraw-Hill)}
\ref{[14] Gell-Mann M \& Hartle J B 1993}\PR{D47}{3345}
\endgroup
\bye